# Tying Research Funding to Progress on Inclusion[1]

Lead:
Dara Norman, NOAO, dnorman@noao.edu

Co-Is:
Terri Brandt, NASA/GSFC, t.j.brandt@nasa.gov,
Zack Berta-Thompson, U of Colorado, zach.bertathompson@colorado.edu,
Natalia Lewandowsha, Green Bank Observatory, nlewando@nrao.edu,
Karen Knierman, Arizona State Univ, karen.knierman@asu.edu,
Nancy Chanover, New Mexico State University, nchanove@nmsu.edu,
Jena Whitley, U Texas-Arlington, jenawhitley@gmail.com,
Aparna Venkatesa, University of San Francisco, avenkatesan@usfca.edu,
Kim Coble, San Francisco State University, kcoble@sfsu.edu,
Adam Burgasser University of California, San Diego aburgasser@gmail.com,
Marie Lemoine-Busserolle, Gemini Observatory, mbussero@gemini.edu,
Kelle Cruz, CUNY Hunter College, kellecruz@gmail.com,
Christopher S. Moore, Harvard-Smithsonian Center for Astrophysics, christopher.s.moore@cfa.harvard.edu,
Kwame Osei-Sarfo, Columbia University, ko2439@columbia.edu,
Sheila Kannappan, UNC Chapel Hill, sheila@physics.unc.edu

Endorsers:
Lia Corrales, University of Michigan, liac@umich.edu,
Alexander L. Rudolph, California State Polytechnic University, alrudolph@cpp.edu,
Alyson Brooks, Rutgers, the State University of New Jersey, abrooks@physics.rutgers.edu,
Kevin Covey, Western Washington University, coveyk@wwu.edu

Type of Activity: State of the Profession; Other: Funding for Projects/Activities

---

[1] The views expressed represent the personal views of the authors, contributors, and endorsers and not necessarily the views of their affiliated institutions.

**Summary of Key Issue and Impact on the Field**

The US professional astronomy and astrophysics fields are not representative of the diversity of people in the nation. For example, 2017 AIP reports show that in 2014, women made up only about 20% of the faculty in astronomy and physics departments, and the numbers for under-represented minorities (men and women) were, and remain, low. However numerous studies have demonstrated that diverse groups (in both cognition and identity) outperform groups that are more homogeneous, even when the homogeneous group is comprised of all 'high achieving experts.' (Hong and Page, 2004, Kleinberg and Raghu, 2018). This has been shown to be the case on a variety of complex tasks. Thus, if we want the best opportunity to make progress on and answer the research questions of the 2020s, we must employ diverse teams who bring different heuristics and perspectives to those problems.

However, currently in the field there are few tangible motivations to encourage projects, missions or programs to employ teams that are diverse in both cognitive areas and identity to take on these complex problems. Managing groups and organizations contracted to run these efforts are currently not required or incentivized to employ an identity diverse workforce.
In this position (white) paper, *we recommend that agency funding (from NSF, NASA, DOE, etc.), especially for missions, projects and programs, encourage the development and retention of diverse teams by requiring documentation of and progress on metrics related to diversity, inclusion and equity. We further recommend that documented progress on diversity and inclusion metrics should be monitored in reviews alongside project management and budget reporting. Managing groups and organizations proposing to administer projects on behalf of agencies should be required to demonstrate competency with respect to diversity and inclusion metrics.*

**Complex scientific problems in the next decade require diverse teams**

There is no doubt that in the next decade astronomy and astrophysics (hereafter astronomy) will be moving forward towards exploring many complex scientific questions about the structure and evolution of the universe, understanding Black Holes and the search for and characterization of Earth-like planets - to name but a few. To address many of these complex questions, improved software, analysis, methods, instrumentation and telescopes will need to be developed, built and implemented.[2] The infrastructure to support the best science will necessarily require diverse teams with expertise in a variety of areas. However, studies also show that it is not only traditionally recognized expertise that is important to teams solving complex problems, but equally important are the tools, strategies and perspectives that team members bring to the problems because of their **identity diversity** and their life experiences resulting from their identities, including gender, race and ethnicity. (Kleinberg and Raghu, 2018).

There is substantial literature on both how and why diversity in groups is key to solving the most difficult and complex problems. Hong and Page (2004) point out that, "groups of diverse problem solvers can outperform groups of high-ability problem solvers". This follows from the fact that high-ability homogeneous problem solvers often bring similar heuristics to finding solutions. However, teams of people from more diverse backgrounds, abilities and perspectives are better able to bring different tools to problem solving and thus identify better solutions. Other work points out that the needed diversity to reap these benefits is not only cognitive, but also depends on identity diversity, that is, the perspective one has because of who they are (i.e., characteristics like their gender, race, ethnicity, socio-economic background etc.) Work by Phillips, K. (2014), Freeman & Huang (2014), and others suggests that in these more diverse groups, all members work harder and pay more attention to aspects of the problem as they work together on solutions, perhaps because of anticipating challenges.

Large and complex projects, missions and collaborations are becoming the norm in astronomy. These programs often require some amount of public funding from federal agencies in order to accomplish their science goals. Requests for public money, for example from the NSF, currently come with a requirement for proposals to provide a description of the project's "Broader Impacts". The definition of 'Broader Impacts' and how a group chooses to fulfil the requirement is left open for arguments to be made about what satisfies this obligation. Often these 'Broader Impacts' sections have been focused largely on education and public outreach activities or technology development. Many in the reviewing community have come to think of 'Broader Impacts' as sections to describe these limited impact areas.   However, expanding the diversity and inclusivity of the astronomy enterprise workforce would have significant impacts in the larger society as well as advancing the growth and progress of the field.

---

[2] For examples see Astro2020 Science White Papers: K. Olsen, et al. 'Science Platforms for Resolved StellarPopulations in the Next Decade'; M. Ntampaka, et al., 'The Role of Machine Learning in the Next Decade of Cosmology'; P. Chang et al., 'Cyberinfrastructure Requirements to EnhanceMulti-messenger Astrophysics'; G. Fabbiano, et al., 'Increasing the Discovery Space in Astrophysics: The Exploration Question for Planetary Systems'; A. Siemiginowska, et al.,' The Next Decade of Astroinformatics and Astrostatistics'.

 The 'astronomy enterprise' includes scientific research as well as the technical and infrastructure development that support research (e.g., instrument building, software development, etc.).  As more public resources are needed to support both the building and operation of  large ground-based projects and space missions, additional requirements for strengthening a more diverse astronomy workforce should be part of the scope of broader impacts to maximize the lasting benefits of the public funding investment.

**Community recognition that diverse teams benefit the astronomy enterprise**

Currently in the astronomy field, there are a number of efforts to expand diversity, inclusion and equity in order to achieve the highest goals and the best scientific outcomes. For example, there are a number of bridge programs that have helped to promote and retain many more URM[3] to obtain PhDs in the field of astronomy.  For example, over the last four years, 27 of 33 (82%) Cal-Bridge Scholars have begun or will be attending PhD programs in physics or astronomy at top PhD programs nationally. [4]  The Columbia University Bridge Program also boosts a high completion rates with 100% of the scholars from the most recent four cohorts admitted into graduate programs (in several fields including astronomy). To date, 11 alumni/ae of the Columbia Bridge Program have received their doctorates with the expectation that at least five additional scholars will receive their doctorates by the end of 2020.[5]

Conferences like Inclusive Astronomy 2015[6] and the Women in Astronomy series (I-IV)[7] have attracted many members of the community to discuss better practices for restructuring the field in ways that are, not only more diverse, inclusive and equitable, but also more welcoming for researchers and students of all backgrounds and abilities. The Sloan Digital Sky Survey (SDSS) collaboration even has a dedicated internal committee, the 'Committee On Inclusion iN SDSS (COINS)'[8], that surveys the climate of diversity and inclusion within the project and reports on progress and concerns to the project leadership.

---

[3] Under-represented minorities
[4] Private communication with Alexander L. Rudolph Director, Cal-Bridge and  CAMPARE/CHAM, California State Polytechnic University, see https://www.cpp.edu/~sci/physics-astronomy/research/cal-bridge.shtml
[5] Private communication with Kwame Osei-Sarfo, Director Bridge to the Ph.D. Program in STEM, Columbia University, see https://bridgetophd.facultydiversity.columbia.edu
[6] https://aas.org/media/press-releases/aas-endorses-vision-statement-inclusive-astronomy
[7] http://www.cvent.com/events/women-in-astronomy-iv-the-many-faces-of-women-astronomers/event-summary-589214b84ab94f26ac269ad9823ef977.aspx

There has also been a trend for the selection committees of oversubscribed astronomy conferences, workshops and schools to actively select for a diverse cohort of attendees. Many of these meetings and groups, including dotAstronomy[9] and the Data Science Fellowship Program[10], employ software, like Entrofy[11], to help them maximize diversity along a number of cognitive and identity parameters. These conferences and programs have been held up in many arenas as models for harnessing creative thought and information exchange in the community.

These efforts clearly demonstrate that the astronomy community is interested and invested in the promotion of diversity to achieve science goals. These many efforts to build diversity into teams of students and researchers are being led on relatively small scales by dedicated individuals to support relatively few students and researchers. Their efforts have been extraordinary for beginning the process of increasing diversity and inclusion in the field, but these efforts are not 'trickling up' to large scale missions, projects and programs. Galinsky et al. (2015) point out that while empirical evidence demonstrates that diversity creates and sustains economic growth, improves decision making, and produces new innovations there is "other research that has identified barriers that limit current diversity levels and produce psychological resistance to efforts to increase diversity." They suggest that policies are therefore needed to promote the diversity present in groups, communities, and nations. Thus, within astronomy, ***funding agencies need to be proactive to promote more diversity of teams in these federally-funded efforts.***

**Funding Agencies Can Support Diverse Teams Through Leadership**

In a recent Facebook post, Thomas Zurbruchen, Associate Administrator for NASA's Science Mission Directorate, recognizes that his own experience has shown him that "Groupthink is the enemy of excellence." He points out that even in high-level NASA meetings, "Groupthink - as is common in teams without sufficient diversity of thought - can creep up in any place and at any time if a community is not deliberate about actively seeking and including diverse viewpoints, and fostering a community of "different" not a community of "same". Running diverse teams needs to be focused on and learned [from]!"  ***We, therefore, recommend that agencies require a review of diversity and inclusion metrics for missions and projects as another axis of excellence on which these programs should be evaluated and held accountable, in addition to science goals.  Diversity and inclusion metrics should be required and monitored in readiness, and other, reviews alongside project management and budget planning.***

---

[8] https://www.sdss.org/collaboration/coins/
[9] https://www.dotastronomy.com/
[10] https://astrodatascience.org/
[11] https://github.com/dhuppenkothen/entrofy

We suggest that science missions, projects, programs, or organizations proposing for funding from government agencies, be required to demonstrate a plan to diversify their scientific and technical workforce, at all levels, as an integral part of the project. Furthermore, implementation of this plan should be part of the evaluation and reporting process to receive and maintain funding. There are a number of metrics that can be assessed to monitor and hold accountable a proposer's commitment to and progress towards inclusion and diversity within their funded project.

Some examples of metrics by which proposals can be reviewed are:
- Establishing an organized (sub)committee that promotes and reports on diversity and inclusion in research activities to the project leadership (e.g. SDSS' COINS);
- Providing evidence of acting on the recommendations of such committees;
- Providing training and presentation opportunities for team members from under-represnted groups;
- Promoting the legacy value of data taken, products produced, codes written, and information shared;
- Planning for and implementing broad and easy access to data and other legacy resources, especially through methods that support inclusion by underserved groups and institutions (e.g., NOAO's Data Lab, and other science platforms);
- Promoting opportunities for "Open Collaboration",  especially to researchers from underserved groups and institutions (e.g., LSST science collaborations);
- Demonstrating that collaboration opportunities result in demographic diversity improvement;
- Promoting opportunities for equitable collaboration;
- Establishing conduct codes and inclusive best practices for communication and interaction;
- Establishing a scientific and administrative governance that is inclusive and transparent;
- Monitoring the demographic information of mission/project/collaboration participants;

- Monitoring the demographic make-up of those in leadership roles.

These are just some examples of diversity and inclusion metrics. Proposers should also be encouraged to identify and implement other metrics that best help them to build diverse teams. However, in many cases this also means that proposers should be required to demonstrate competency on the topics of diversity metrics. Like with scientific goals, identification of and experience with proper metrics on diversity requires knowledge and expertise in these areas. Review of proposed metrics will need to be assessed as achievable by reviewers expert and experienced in evaluating areas of diversity and inclusion.

S.Page (2017) suggests several points to keep in mind when implementing or designing metrics to decide on the success of implementing diverse teams. (Also see details in his talks, e.g., https://imdb.yt/watch/scott-page-talk-diversity-productivity-10425527 ). First heuristic difference is not embedded in DNA, it is encompassed in one's experience, thus just assembling teams based on demographics alone will not provide the bonus that diversity has the potential to bring. Secondly, any diversity boost to problem solving depends critically on the extent to which there is collaboration, communication, and trust within teams. Proper management of a diverse team is critical and its importance, and requirement of resources, should not be underestimated. Another place where management becomes important is in the culture of the group, thus leadership should encourage constructive disagreement and dissent in order to avoid 'groupthink'. This is where metrics like ethics codes and best practices for communication become important so that trust is established and allows for group members to be honest with and respectful of one another. Finally, to properly leverage team diversity, leaders must understand the value of the solution that the group is trying to identify and thus the value of having different heuristics and perspectives being part of that solution. Leveraging diversity often requires restructuring current antiquated ways of establishing teams and their dynamics.

**Concluding Remarks**
The decadal survey review is about identifying where the field of astronomy is now and predicting where the science will go in the future. It is about prioritizing the field's resources to take advantage of those near- and long-term science goals. A key part is promoting the best workforce, opportunities and infrastructure for success in reaching those goals. The evidence has shown that moving beyond good ideas to groundbreaking ideas must include diverse teams. However, currently the field is significantly lacking in its diversity and unable to effectively benefit from what diversity there is.  Federal agencies must play an active role in promoting the inclusion of diverse teams and leading the way to the most innovative research in the 2020s.